\documentclass[twoside,a4paper]{amsart}
\providecommand{\journal}[1]{}
\newenvironment{frontmatter}{}{\maketitle}
\providecommand{\sep}{ }
\providecommand{\MSC}[1][2008]{MSC#1: }

\usepackage[breakable]{tcolorbox}
\usepackage{graphicx}

\usepackage{xcolor} 

\definecolor{incolor}{HTML}{303F9F}
\definecolor{outcolor}{HTML}{D84315}
\definecolor{cellborder}{HTML}{CFCFCF}
\definecolor{cellbackground}{HTML}{F7F7F7}

    \AtBeginDocument{%
    }
\usepackage{fancyvrb} 

    \makeatletter
    \newcommand{\boxspacing}{\kern\kvtcb@left@rule\kern\kvtcb@boxsep}
    \makeatother
    \newcommand{\prompt}[4]{
        \ttfamily\llap{{\color{#2}[#3]:\hspace{3pt}#4}}\vspace{-\baselineskip}
    }

\makeatletter
\def\PY@reset{\let\PY@it=\relax \let\PY@bf=\relax%
    \let\PY@ul=\relax \let\PY@tc=\relax%
    \let\PY@bc=\relax \let\PY@ff=\relax}
\def\PY@tok#1{\csname PY@tok@#1\endcsname}
\def\PY@toks#1+{\ifx\relax#1\empty\else%
    \PY@tok{#1}\expandafter\PY@toks\fi}
\def\PY@do#1{\PY@bc{\PY@tc{\PY@ul{%
    \PY@it{\PY@bf{\PY@ff{#1}}}}}}}
\def\PY#1#2{\PY@reset\PY@toks#1+\relax+\PY@do{#2}}

\expandafter\def\csname PY@tok@w\endcsname{\def\PY@tc##1{\textcolor[rgb]{0.73,0.73,0.73}{##1}}}
\expandafter\def\csname PY@tok@c\endcsname{\let\PY@it=\textit\def\PY@tc##1{\textcolor[rgb]{0.25,0.50,0.50}{##1}}}
\expandafter\def\csname PY@tok@cp\endcsname{\def\PY@tc##1{\textcolor[rgb]{0.74,0.48,0.00}{##1}}}
\expandafter\def\csname PY@tok@k\endcsname{\let\PY@bf=\textbf\def\PY@tc##1{\textcolor[rgb]{0.00,0.50,0.00}{##1}}}
\expandafter\def\csname PY@tok@kp\endcsname{\def\PY@tc##1{\textcolor[rgb]{0.00,0.50,0.00}{##1}}}
\expandafter\def\csname PY@tok@kt\endcsname{\def\PY@tc##1{\textcolor[rgb]{0.69,0.00,0.25}{##1}}}
\expandafter\def\csname PY@tok@o\endcsname{\def\PY@tc##1{\textcolor[rgb]{0.40,0.40,0.40}{##1}}}
\expandafter\def\csname PY@tok@ow\endcsname{\let\PY@bf=\textbf\def\PY@tc##1{\textcolor[rgb]{0.67,0.13,1.00}{##1}}}
\expandafter\def\csname PY@tok@nb\endcsname{\def\PY@tc##1{\textcolor[rgb]{0.00,0.50,0.00}{##1}}}
\expandafter\def\csname PY@tok@nf\endcsname{\def\PY@tc##1{\textcolor[rgb]{0.00,0.00,1.00}{##1}}}
\expandafter\def\csname PY@tok@nc\endcsname{\let\PY@bf=\textbf\def\PY@tc##1{\textcolor[rgb]{0.00,0.00,1.00}{##1}}}
\expandafter\def\csname PY@tok@nn\endcsname{\let\PY@bf=\textbf\def\PY@tc##1{\textcolor[rgb]{0.00,0.00,1.00}{##1}}}
\expandafter\def\csname PY@tok@ne\endcsname{\let\PY@bf=\textbf\def\PY@tc##1{\textcolor[rgb]{0.82,0.25,0.23}{##1}}}
\expandafter\def\csname PY@tok@nv\endcsname{\def\PY@tc##1{\textcolor[rgb]{0.10,0.09,0.49}{##1}}}
\expandafter\def\csname PY@tok@no\endcsname{\def\PY@tc##1{\textcolor[rgb]{0.53,0.00,0.00}{##1}}}
\expandafter\def\csname PY@tok@nl\endcsname{\def\PY@tc##1{\textcolor[rgb]{0.63,0.63,0.00}{##1}}}
\expandafter\def\csname PY@tok@ni\endcsname{\let\PY@bf=\textbf\def\PY@tc##1{\textcolor[rgb]{0.60,0.60,0.60}{##1}}}
\expandafter\def\csname PY@tok@na\endcsname{\def\PY@tc##1{\textcolor[rgb]{0.49,0.56,0.16}{##1}}}
\expandafter\def\csname PY@tok@nt\endcsname{\let\PY@bf=\textbf\def\PY@tc##1{\textcolor[rgb]{0.00,0.50,0.00}{##1}}}
\expandafter\def\csname PY@tok@nd\endcsname{\def\PY@tc##1{\textcolor[rgb]{0.67,0.13,1.00}{##1}}}
\expandafter\def\csname PY@tok@s\endcsname{\def\PY@tc##1{\textcolor[rgb]{0.73,0.13,0.13}{##1}}}
\expandafter\def\csname PY@tok@sd\endcsname{\let\PY@it=\textit\def\PY@tc##1{\textcolor[rgb]{0.73,0.13,0.13}{##1}}}
\expandafter\def\csname PY@tok@si\endcsname{\let\PY@bf=\textbf\def\PY@tc##1{\textcolor[rgb]{0.73,0.40,0.53}{##1}}}
\expandafter\def\csname PY@tok@se\endcsname{\let\PY@bf=\textbf\def\PY@tc##1{\textcolor[rgb]{0.73,0.40,0.13}{##1}}}
\expandafter\def\csname PY@tok@sr\endcsname{\def\PY@tc##1{\textcolor[rgb]{0.73,0.40,0.53}{##1}}}
\expandafter\def\csname PY@tok@ss\endcsname{\def\PY@tc##1{\textcolor[rgb]{0.10,0.09,0.49}{##1}}}
\expandafter\def\csname PY@tok@sx\endcsname{\def\PY@tc##1{\textcolor[rgb]{0.00,0.50,0.00}{##1}}}
\expandafter\def\csname PY@tok@m\endcsname{\def\PY@tc##1{\textcolor[rgb]{0.40,0.40,0.40}{##1}}}
\expandafter\def\csname PY@tok@gh\endcsname{\let\PY@bf=\textbf\def\PY@tc##1{\textcolor[rgb]{0.00,0.00,0.50}{##1}}}
\expandafter\def\csname PY@tok@gu\endcsname{\let\PY@bf=\textbf\def\PY@tc##1{\textcolor[rgb]{0.50,0.00,0.50}{##1}}}
\expandafter\def\csname PY@tok@gd\endcsname{\def\PY@tc##1{\textcolor[rgb]{0.63,0.00,0.00}{##1}}}
\expandafter\def\csname PY@tok@gi\endcsname{\def\PY@tc##1{\textcolor[rgb]{0.00,0.63,0.00}{##1}}}
\expandafter\def\csname PY@tok@gr\endcsname{\def\PY@tc##1{\textcolor[rgb]{1.00,0.00,0.00}{##1}}}
\expandafter\def\csname PY@tok@ge\endcsname{\let\PY@it=\textit}
\expandafter\def\csname PY@tok@gs\endcsname{\let\PY@bf=\textbf}
\expandafter\def\csname PY@tok@gp\endcsname{\let\PY@bf=\textbf\def\PY@tc##1{\textcolor[rgb]{0.00,0.00,0.50}{##1}}}
\expandafter\def\csname PY@tok@go\endcsname{\def\PY@tc##1{\textcolor[rgb]{0.53,0.53,0.53}{##1}}}
\expandafter\def\csname PY@tok@gt\endcsname{\def\PY@tc##1{\textcolor[rgb]{0.00,0.27,0.87}{##1}}}
\expandafter\def\csname PY@tok@err\endcsname{\def\PY@bc##1{\setlength{\fboxsep}{0pt}\fcolorbox[rgb]{1.00,0.00,0.00}{1,1,1}{\strut ##1}}}
\expandafter\def\csname PY@tok@kc\endcsname{\let\PY@bf=\textbf\def\PY@tc##1{\textcolor[rgb]{0.00,0.50,0.00}{##1}}}
\expandafter\def\csname PY@tok@kd\endcsname{\let\PY@bf=\textbf\def\PY@tc##1{\textcolor[rgb]{0.00,0.50,0.00}{##1}}}
\expandafter\def\csname PY@tok@kn\endcsname{\let\PY@bf=\textbf\def\PY@tc##1{\textcolor[rgb]{0.00,0.50,0.00}{##1}}}
\expandafter\def\csname PY@tok@kr\endcsname{\let\PY@bf=\textbf\def\PY@tc##1{\textcolor[rgb]{0.00,0.50,0.00}{##1}}}
\expandafter\def\csname PY@tok@bp\endcsname{\def\PY@tc##1{\textcolor[rgb]{0.00,0.50,0.00}{##1}}}
\expandafter\def\csname PY@tok@fm\endcsname{\def\PY@tc##1{\textcolor[rgb]{0.00,0.00,1.00}{##1}}}
\expandafter\def\csname PY@tok@vc\endcsname{\def\PY@tc##1{\textcolor[rgb]{0.10,0.09,0.49}{##1}}}
\expandafter\def\csname PY@tok@vg\endcsname{\def\PY@tc##1{\textcolor[rgb]{0.10,0.09,0.49}{##1}}}
\expandafter\def\csname PY@tok@vi\endcsname{\def\PY@tc##1{\textcolor[rgb]{0.10,0.09,0.49}{##1}}}
\expandafter\def\csname PY@tok@vm\endcsname{\def\PY@tc##1{\textcolor[rgb]{0.10,0.09,0.49}{##1}}}
\expandafter\def\csname PY@tok@sa\endcsname{\def\PY@tc##1{\textcolor[rgb]{0.73,0.13,0.13}{##1}}}
\expandafter\def\csname PY@tok@sb\endcsname{\def\PY@tc##1{\textcolor[rgb]{0.73,0.13,0.13}{##1}}}
\expandafter\def\csname PY@tok@sc\endcsname{\def\PY@tc##1{\textcolor[rgb]{0.73,0.13,0.13}{##1}}}
\expandafter\def\csname PY@tok@dl\endcsname{\def\PY@tc##1{\textcolor[rgb]{0.73,0.13,0.13}{##1}}}
\expandafter\def\csname PY@tok@s2\endcsname{\def\PY@tc##1{\textcolor[rgb]{0.73,0.13,0.13}{##1}}}
\expandafter\def\csname PY@tok@sh\endcsname{\def\PY@tc##1{\textcolor[rgb]{0.73,0.13,0.13}{##1}}}
\expandafter\def\csname PY@tok@s1\endcsname{\def\PY@tc##1{\textcolor[rgb]{0.73,0.13,0.13}{##1}}}
\expandafter\def\csname PY@tok@mb\endcsname{\def\PY@tc##1{\textcolor[rgb]{0.40,0.40,0.40}{##1}}}
\expandafter\def\csname PY@tok@mf\endcsname{\def\PY@tc##1{\textcolor[rgb]{0.40,0.40,0.40}{##1}}}
\expandafter\def\csname PY@tok@mh\endcsname{\def\PY@tc##1{\textcolor[rgb]{0.40,0.40,0.40}{##1}}}
\expandafter\def\csname PY@tok@mi\endcsname{\def\PY@tc##1{\textcolor[rgb]{0.40,0.40,0.40}{##1}}}
\expandafter\def\csname PY@tok@il\endcsname{\def\PY@tc##1{\textcolor[rgb]{0.40,0.40,0.40}{##1}}}
\expandafter\def\csname PY@tok@mo\endcsname{\def\PY@tc##1{\textcolor[rgb]{0.40,0.40,0.40}{##1}}}
\expandafter\def\csname PY@tok@ch\endcsname{\let\PY@it=\textit\def\PY@tc##1{\textcolor[rgb]{0.25,0.50,0.50}{##1}}}
\expandafter\def\csname PY@tok@cm\endcsname{\let\PY@it=\textit\def\PY@tc##1{\textcolor[rgb]{0.25,0.50,0.50}{##1}}}
\expandafter\def\csname PY@tok@cpf\endcsname{\let\PY@it=\textit\def\PY@tc##1{\textcolor[rgb]{0.25,0.50,0.50}{##1}}}
\expandafter\def\csname PY@tok@c1\endcsname{\let\PY@it=\textit\def\PY@tc##1{\textcolor[rgb]{0.25,0.50,0.50}{##1}}}
\expandafter\def\csname PY@tok@cs\endcsname{\let\PY@it=\textit\def\PY@tc##1{\textcolor[rgb]{0.25,0.50,0.50}{##1}}}

\def\PYZbs{\char`\\}
\def\PYZus{\char`\_}

\def\PYZpc{\char`\%}

\def\PYZhy{\char`\-}

\def\PYZdq{\char`\"}

\makeatother
    
\usepackage{amssymb}
\usepackage{amsthm,amsmath}

\usepackage{listings}
\usepackage{lineno}
\lstdefinestyle{Python}{
    language        = Python,
    basicstyle      = \sffamily,
    keywordstyle    = \bfseries,
    stringstyle     = \rmfamily,
    commentstyle    = \sffamily
}

\journal{Name of journal}

\usepackage{hyperref}

\providecommand{\cites}[1]{\cite{#1}}
\providecommand{\Space}[3][]{\ensuremath{\mathbb{#2}^{#3}_{#1}{}}}

\providecommand{\Cliff}[2][\comment]{{\ensuremath{%
\mathcal{C}\kern-0.18em\ell(#1,#2)}}}
\providecommand{\norm}[2][\relax]{\left\|#2\right\|\ifx#1\relax\else_{#1}\fi}
\providecommand{\modulus}[2][\relax]{\left| #2 \right|\ifx#1\relax\else_{#1}\fi}

\providecommand{\scalar}[3][\relax]{\left\langle #2,#3
        \right\rangle\ifx#1\relax\else_{#1}\fi}
\providecommand{\nscalar}[3][\relax]{\left[ #2,#3
        \right]\ifx#1\relax\else_{#1}\fi}
\providecommand{\wiki}[2]{\href{http://en.wikipedia.org/wiki/#1}{#2}}
\providecommand{\Zbl}[1]{Zbl\href{http://www.emis.de:80/cgi-bin/zmen/ZMATH/en/zmathf.html?first=1&maxdocs=3&type=html&an=#1&format=complete}{#1}}

\newcommand*\vtick{\kern -.1em\textsc{\char13}}

\newcommand{\CPP}{\textsf{C\nolinebreak\hspace{-.05em}\raisebox{.4ex}{\tiny\bf +}\nolinebreak\hspace{-.10em}\raisebox{.4ex}{\tiny\bf +}}}

\newcommand{\Python}{\textsf{Python}}
\newcommand{\Jupyter}{\textsf{Jupyter}}
\newcommand{\NoWEB}{\texttt{noweb}}
\providecommand{\MetaPost}{\texttt{Meta}\-\texttt{Post}}
\providecommand{\GiNaC}{\textsf{GiNaC}}
\providecommand{\MoebInv}{\textsf{MoebInv}}
\providecommand{\pyGiNaC}{\textsf{pyGiNaC}}
\providecommand{\cycle}[3][]{{#1 C^{#2}_{#3}}}

\providecommand{\Asymptote}{\texttt{Asymptote}}
\providecommand{\myeprint}[2]{E-print: \href{#1}{\texttt{#2}}}

\newif\iftth

\providecommand{\clifford}[2][]{\ifcase #1 #2\or \tilde{#2} \or \breve{#2} \fi}
  
  \theoremstyle{definition}

  \theoremstyle{remark}
  
\DeclareFontFamily{OT1}{cyr}{}
\DeclareFontShape{OT1}{cyr}{m}{n}
   {  <5> <6> <7> <8> <9> gen * wncyr
      <10> <10.95> <12> <14.4> <17.28> <20.74> <24.88> wncyr10}{}
\DeclareFontShape{OT1}{cyr}{m}{it}
    {
       <5> <6> <7> <8> <9> gen * wncyi
      <10> <10.95> <12> <14.4> <17.28> <20.74> <24.88>wncyi10
      }{}
\DeclareFontShape{OT1}{cyr}{m}{ss}
    {
       <5> <6> <7> <8> wncyss8
       <9> wncy9
      <10> <10.95> <12> <14.4> <17.28> <20.74> <24.88>wncyss10
      }{}
\DeclareFontShape{OT1}{cyr}{m}{sc}
    {
       <5> <6> <7> <8> <9> <10> <10.95> <12> <14.4> <17.28> <20.74> <24.88>wncysc10
      }{}
\DeclareFontShape{OT1}{cyr}{bx}{n}
   {
       <5> <6> <7> <8> <9> gen * wncyb
      <10> <10.95> <12> <14.4> <17.28> <20.74> <24.88>wncyb10
      }{}
\DeclareTextFontCommand{\textcyr}{\fontfamily{cyr}\selectfont}
\providecommand{\cyr}{\fontfamily{cyr}\selectfont\def\cprime{\~}}
\providecommand{\cprime}{\vtick}

        \providecommand{\MR}[1]{MR\href{http://www.ams.org/mathscinet-getitem?mr=#1}{#1}}
  \providecommand{\Zbl}[1]{Zbl\href{http://www.emis.de:80/cgi-bin/zmen/ZMATH/en/zmathf.html?first=1&maxdocs=3&type=html&an=#1&format=complete}{#1}}

  \providecommand{\myeprint}[2]{E-print: \href{#1}{\texttt{#2}}}
  \providecommand{\amscite}[3]{\cite[#3]{#1}}
\begin{document}

\begin{frontmatter}


\title{MoebInv: C++ libraries for manipulations in non-Euclidean geometry}


\author{Vladimir V. Kisil}

\address{School of Mathematics,
University of Leeds,
Leeds LS2\,9JT,
England}

\begin{abstract}
  The introduced package \MoebInv\ contains two \CPP\ libraries for
  symbolic, numeric and graphical manipulations in non-Euclidean
  geometry.  The first library \texttt{cycle} implements basic
  geometric operations on cycles, which are the zero sets of certain
  polynomials of degree two. The second library \texttt{figure}
  operates on ensembles of cycles interconnected by Moebius-invariant
  relations: orthogonality, tangency, etc.  Both libraries work in
  spaces with any dimension and arbitrary signatures of their metrics.
%
  Their essential functionality is accessible in
  interactive modes from \Python/\Jupyter\ shells and a dedicated
  Graphical User Interface. The latter does not require any coding
  skills and can be used in education.
%
%
  The package is tested on (and supplied for) various Linux
  distributions, Windows\,10, Mac\,OS\,X and several cloud services.
\end{abstract}

\begin{keyword}{
  Lie spheres geometry \sep
  M\"obius transformations \sep
  fractional linear transformations \sep
  symbolic computations
  \MSC[2008]
68U05 \sep 51B25  \sep 68W30 \sep 51N25 \sep 51B10 \sep 11E88}
\end{keyword}

\end{frontmatter}




\section{Introduction}
\label{sec:introduction}

We present the Open Source package
\MoebInv~\cites{Kisil05b,Kisil14b}---a research and educational tool
for various geometric setups. Its domain, design and functionality
have some unique features which are not available elsewhere. The code
is symbiotically growing together with the research in the extended
M\"obius--Lie geometry~\cite{Kisil19a}, both---the code and the
theory---benefited from this interaction.  Functionality of the
package is accessible from a \CPP\ code and can be interactively used
through \Python/\Jupyter\ shells and a dedicated Graphical User
Interface (GUI).

There is already a collection of well-established and reputable Open
Source geometry software (\textsf{GeoGebra}~\cite{GeoGebra},
\textsf{CaRMetal}~\cite{CaRMetal}, \textsf{Kig}~\cite{Kig},
\textsf{Dr.~Geo}~\cite{DrGeo}) as well as commercial educational
packages (The Geometer's Sketchpad, Cabri, Cinderella,
\href{https://www.netpad.net.cn/}{NetPad}).  All of those are designed
to work primary with the Euclidean geometry---the oldest archetypal
mathematical theory. Nowadays it is the first from a large
family of various geometries: affine, projective, conformal,
Riemannian, etc. According to F.~Klein's
\emph{\wiki{Erlangen_program}{Erlangen programme}}%
\index{Erlangen programme|indef} (which was influenced by S.~Lie),
a geometry studies invariant properties under a certain transitive%
\index{transitive}%
\index{action!transitive} group action.

The package \MoebInv\ works with invariants of fractional linear
transformations (FLT) which contain M\"obius maps as an important
subset. The implementation admits spaces of any (including symbolic)
dimensionality and arbitrary signatures of metric (covering the
degenerate cases).  The associated geometries span a wide domain,
which includes conformal~\amscite{Simon11a}*{Ch.~9},
hypercomplex~\cite{Yaglom79} and
\href{https://en.wikipedia.org/wiki/Lie_sphere_geometry}{Lie sphere
  geometries}~\cites{Cecil08a,Benz07a,Juhasz18a}.  For these fields
the package facilitates:
\begin{itemize}
\item Experiments and research.
\item Automated theorem proving and symbolic calculation.
\item Visualisation and interactive manipulations.
\item Exact arithmetic and high precision numeric evaluation.
\item Programming extensions of functionality.
\item Educational usage (starting from school and college levels)
  which does not require any coding skills.
\end{itemize}

For {sake of simplicity} we illustrate this paper by the fundamental
and most visual case of fractional linear transformations (FLT) of the
complex plane defined by \(2\times 2\) complex
matrices~\amscite{Simon11a}*{Ch.~9}:
\begin{equation}
  \label{eq:flt-defn}
  \begin{pmatrix}
    a&b\\c&d
  \end{pmatrix}:\  z \mapsto
  \frac{az+b}{cz+d}\,, \qquad \text{where }
  z\in \Space{C}{} \text{ and }\det\begin{pmatrix}
    a&b\\c&d
  \end{pmatrix}\neq 0.
\end{equation}
A FLT-invariant family of objects unifies circles, lines and
points---all together are called \emph{cycles} in this framework. A
FLT-invariant relations between cycles include incidence,
orthogonality, tangency, etc. The efficiency of the package \MoebInv\
is based on the mathematical formalism~\cites{Kisil14b,Kisil12a} which
encodes the variety of different FLT-invariant relations between
cycles in a unified way.

Besides aesthetical value the conformal and M\"obius--Lie geometries
have countless applications ranging from quantum spin
dynamics~\cites{Jacimovic18a},
cosmology~\cite{CartasEscalanteHerreraGonzalez19a} and integrable
systems~\cite{BobenkoSchief18a} to computer-aided
design~\cite{JuttlerMaroscheckKimHong19a} and physiological
models~\cite{Zhang19a}.

\section{Problems and Background}
\label{sec:problems-background}

From the point of view of synthetic Euclidean geometry a cycle (2D
sphere) is the locus of points on a distance \(R\) from a given point
\((x_0,y_0)\). Its analytic description is encoded in a quadratic
equation:
\begin{equation}
  \label{eq:cycle-eq}
  k(x^2+y^2)-2lx-2ny+m=0, 
\end{equation}
where \(l=kx_0\), \(n=ky_0\), and \(m=l^2+n^2-R^2\) for an arbitrary
\(k\neq 0\).  Although it is theoretically sufficient for analytic
solutions of various geometrical problems, practically its
non-linearity produces rapidly increasing complexity of expression in
symbolic computations and growing rounding errors in numeric
evaluations.

It was only relatively recently
realised~\cites{FillmoreSpringer90a,Cnops02a} that the space of all
spheres possesses a FLT-invariant inner product, which naturally
encodes many fundamental geometric relations---the orthogonality to be
the most important among them.  In addition there is significant set
of FLT-invariant geometric relations, e.g. tangency, Steiner power,
which are genially quadratic. Analytic solutions of several
simultaneous quadratic relations may be very involved. Fortunately, it
was observed~\cites{FillmoreSpringer00a,Kisil14b} that there is a
possibility to reduce such problem to a set of \emph{only one}
quadratic relation and several linear ones. Such a set admits an
effective algorithmic solution even in symbolic setup, which is the
backbone of the package \MoebInv.

More specifically, the first library
\texttt{cycle}~\cites{Kisil05b,Kisil12a,Kisil06a} manipulates
individual cycles within the \GiNaC~\cite{GiNaC} computer algebra
system. The mathematical formalism employed in the library
\texttt{cycle} is based on Clifford algebras and the
Fillmore--Springer--Cnops construction (FSCc), which has a long
history, see \amscite{Schwerdtfeger79a}*{\S~1.1},
  \amscite{Cnops02a}*{\S~4.1}, \cite{FillmoreSpringer90a},
  \amscite{Kirillov06}*{\S~4.2}, \cite{Kisil05a},
  \amscite{Kisil12a}*{\S~4.2}. Compared to a plain analytical
treatment~\cites{Pedoe95a,Benz07a}, FSCc is much more efficient and
conceptually coherent in dealing with FLT-invariant properties of
cycles. Correspondingly, the computer code based on FSCc is easy to
write and maintain.  The second library \texttt{figure} manipulates
ensembles of cycles (quadrics) interrelated by certain FLT-invariant
geometric conditions. There are methods to add, modify and delete
elements of the figure as described in the next section.

\section{Software Framework }
\label{}
The library \texttt{figure} implements the \emph{functional programming}
framework---in contrast to \emph{procedural approach} used in popular
software packages like \textsf{GeoGebra}~\cite{GeoGebra},
\textsf{CaRMetal}~\cite{CaRMetal}, \textsf{Kig}~\cite{Kig},
\textsf{Dr.~Geo}~\cite{DrGeo}. The later provides a fixed set of
geometric construction procedures, e.g. ``find the midpoint of an
interval'', ``drop the perpendicular from a point to a line''. In
contrast, all new cycles in \texttt{class figure} are added through a
list of defining relations, which links the new cycle to already
existing ones\footnote{In fact, it is possible and useful to include
  relations of a new cycle to itself as well. For example, points are
  defined by the condition to be self-orthogonal.}. 

\subsection{Software Architecture}
\label{sec:softw-arch}


 Thinking a cycle ensemble as a graph, one can say that the library
\texttt{cycle} deals with individual vertices (cycles), while \texttt{figure}
considers edges (relations between pairs of cycles) and the whole
graph. Intuitively, an interaction with the library \texttt{figure} reminds
compass-and-straightedge constructions, where new lines, points or circles are
added to a drawing one-by-one through relations to already presented
objects (e.g. the line through two points, the intersection point or the
circle with given centre and a point). 
To avoid ``chicken or the egg'' dilemma all cycles are stored in a
hierarchical structure of generations, numbered by integers. The basic
principles are:
\begin{enumerate}
\item Any explicitly defined cycle (i.e., a cycle which is not related to any
  previously known cycle) is placed into generation-0;
\item Any new cycle defined by relations to \emph{previous} cycles
  from generations \(k_1\), \(k_2\), \ldots, \(k_n\) is placed to the
  generation \(k\) calculated as:
  \begin{equation}
    \label{eq:generation-calculation}
    k=\max(k_1,k_2,\ldots,k_n)+1 .
  \end{equation}
  This rule has an exception that a cycle may have a relation to itself,
  e.g. isotropy (self-orthogonality) condition
  \(\scalar{\cycle{}{}}{\cycle{}{}}=0\), which specifies point-like
  cycles.
\end{enumerate}
If the number or nature of conditions is not
sufficient to define the cycle uniquely (up to natural quadratic
multiplicity), then the cycle will depend on a number of free
(symbolic) variables.

\subsection{Software Functionalities}
\label{sec:softw-funct}


Both libraries are capable to work in spaces of any dimensionality and
metrics with an arbitrary signatures: Euclidean, Minkowski and even
degenerate. Parameters of objects can be symbolic or numeric, the
latter admit calculations with exact or approximate arithmetic.
Drawing routines work with any (elliptic, parabolic or hyperbolic)
metric in two dimensions trough \Asymptote~\cite{Asymptote} software.

There is a macro-like tool, which is called \texttt{subfigure}. Such a
\texttt{subfigure} is a \texttt{figure} itself, such that its inner hierarchy of
generations and relations is not visible from the current
\texttt{figure}. Instead, some cycles (of any generations) of the current
\texttt{figure} are used as predefined cycles of generation-0 of
\texttt{subfigure}. Then only one dependent cycle of \texttt{subfigure}, which
is known as \texttt{result}, is returned back to the current \texttt{figure}. The
generation of the \texttt{result} is calculated from generations of input
cycles by the same formula~\eqref{eq:generation-calculation}.

There is a possibility to test certain conditions (e.g., ``are two
cycles orthogonal?'') or measure certain quantities (e.g., ``what is
their intersection angle?'') for already defined cycles. In
particular, such methods can be used to prove geometrical statements
according to the Cartesian programme, that is replacing the synthetic
geometry by purely algebraic manipulations.  Besides \CPP\ libraries
there is a \Python\ wrapper, which can be used in interactive mode. It
is accessible through cloud computing on several
hosts~\cites{Kisil19c,Kisil19d} with sample \Jupyter\ notebooks. There
is a Graphical User Interface (GUI), which allows to create figures by
mouse clicks. \Python/\Jupyter and GUI are bidirectionally integrated
by data exchange routines. There are two---binary and
human-readable/editable--- formats provided for this purpose.

Metrics of the point space and the cycle spaces do not need to
coincide. Similar bi-metric models was used recently to tackle the
quantum Hall effect~\cite{LapaGromovHughes19a}.
                    

\begin{figure}[htbp]
  \centering
  \makebox[0pt][l]{(a)}\includegraphics[width=.48\textwidth]{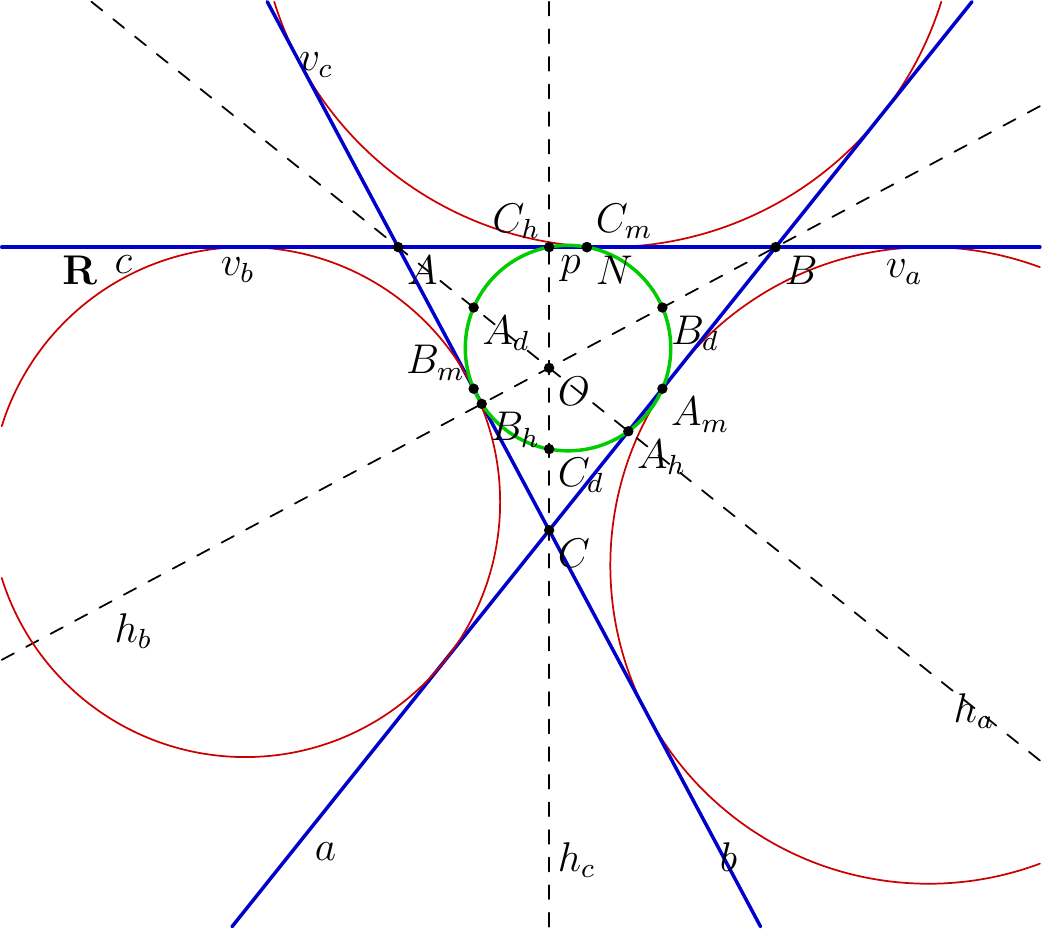}\hfill
  \makebox[0pt][l]{(b)}\includegraphics[width=.48\textwidth]{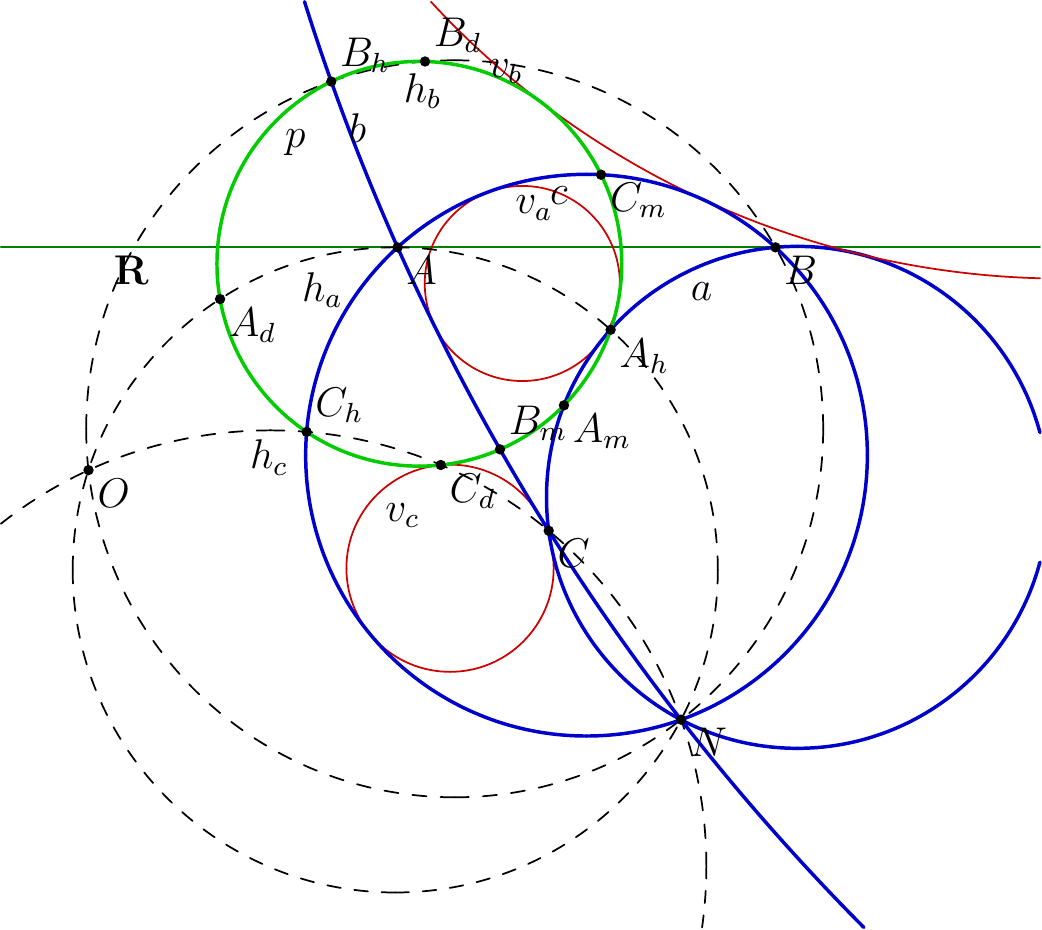}\\[1em]
  \makebox[0pt][l]{(c)}\includegraphics[width=.48\textwidth]{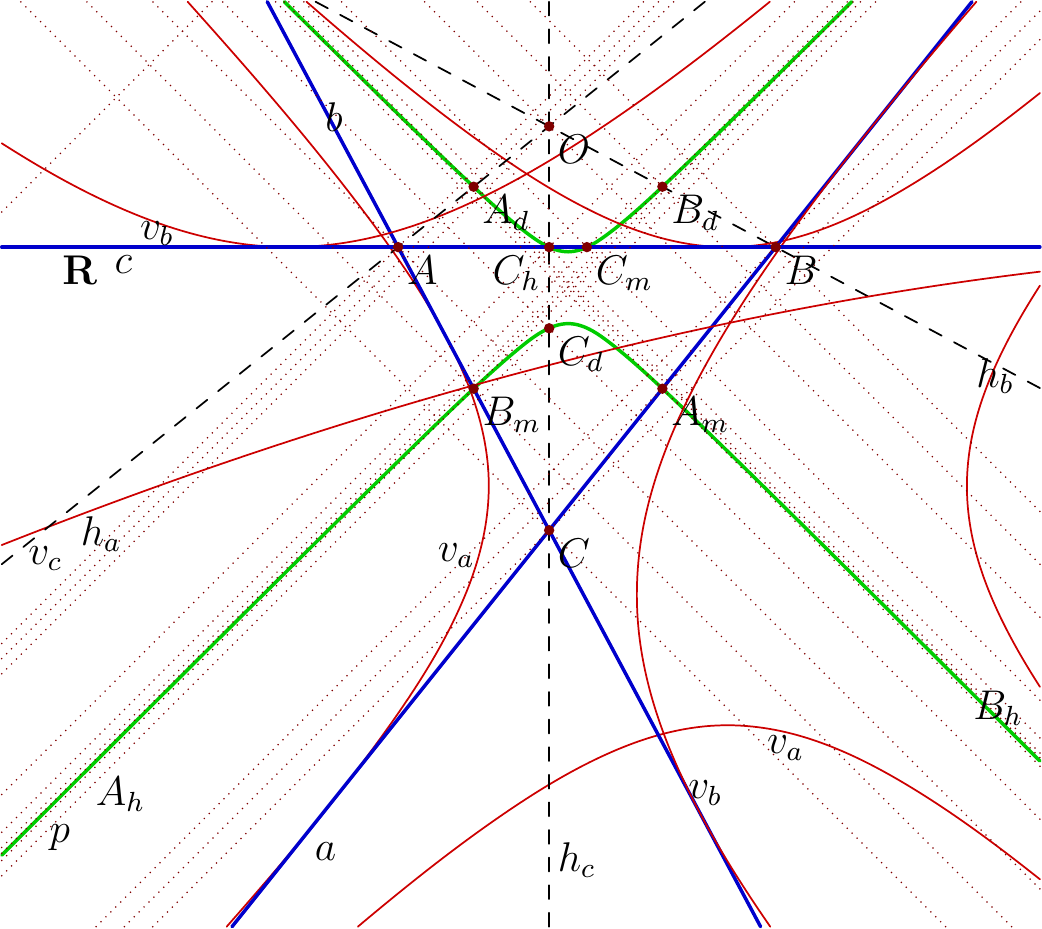}\hfill
  \makebox[0pt][l]{(d)}\includegraphics[width=.48\textwidth]{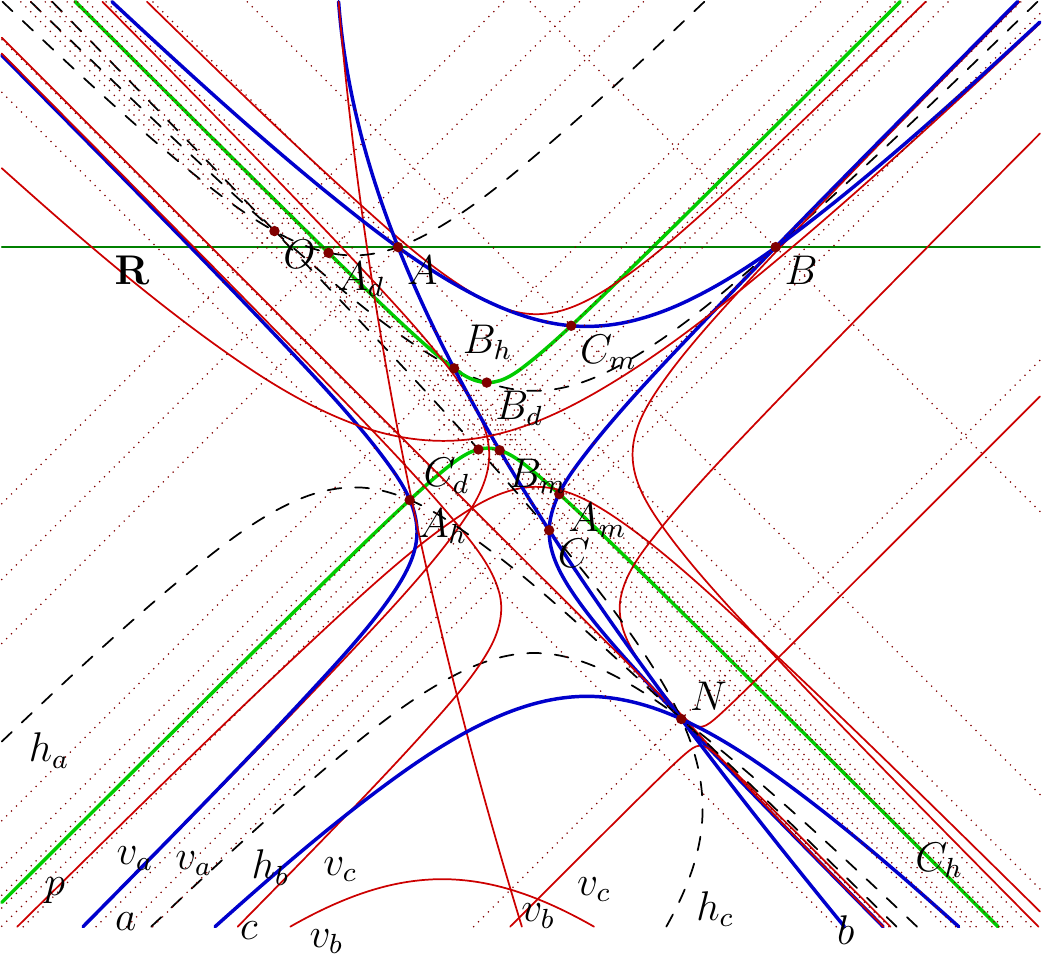}
  \caption[The illustration of the conformal nine-points theorem]
  {The illustration of the conformal nine-points theorem discovered with \MoebInv\ package. The
    left column is the statement for a triangle with straight sides, the right column is its
    conformal version. The
    first row show the elliptic point space, the second row---the
    hyperbolic point space.}
  \label{fig:illustr-conf-nine}
\end{figure}

\section{Illustrative Examples}
\label{sec:illustr-exampl}



As an elementary demonstration, we an analytic proof of a geometrical
statement from \Jupyter\ notebooks\cites{Kisil19c,Kisil19d}. Let \(P\)
be a point of contact of a cycle \(a\) and its tangent \(l\). We will
show that a radius \(r\) of the cycle \(a\) to the point \(P\) is
orthogonal to the line \(l\). To simplify setup we assume that \(a\)
is the unit circle.

    First, we need to load libraries (assuming
\href{https://colab.research.google.com/github/vvkisil/MoebInv-notebooks/blob/master/Introduction/Software_installation_GUI_integration.ipynb}{software
installation} is already done).

    \begin{tcolorbox}[breakable, size=fbox, boxrule=1pt, pad at break*=1mm,colback=cellbackground, colframe=cellborder]
\prompt{In}{incolor}{1}{\boxspacing}
\begin{Verbatim}[commandchars=\\\{\}]
\PY{k+kn}{from} \PY{n+nn}{figure} \PY{k}{import} \PY{o}{*}
\end{Verbatim}
\end{tcolorbox}

    Then, we initialise a figure \(F\) with a default Euclidean metric.

    \begin{tcolorbox}[breakable, size=fbox, boxrule=1pt, pad at break*=1mm,colback=cellbackground, colframe=cellborder]
\prompt{In}{incolor}{2}{\boxspacing}
\begin{Verbatim}[commandchars=\\\{\}]
\PY{n}{F}\PY{o}{=}\PY{n}{figure}\PY{p}{(}\PY{p}{)}
\end{Verbatim}
\end{tcolorbox}

    We add the unit circle \(a\) to the figure by specifying the explicit
coefficients \((1, 0, 0, -1)\) of its quadratic equations, cf.~\eqref{eq:cycle-eq}:
\[1\cdot(x^2+y^2)-0\cdot x -0\cdot y -1 =0\].

    \begin{tcolorbox}[breakable, size=fbox, boxrule=1pt, pad at break*=1mm,colback=cellbackground, colframe=cellborder]
\prompt{In}{incolor}{3}{\boxspacing}
\begin{Verbatim}[commandchars=\\\{\}]
\PY{n}{a}\PY{o}{=}\PY{n}{F}\PY{o}{.}\PY{n}{add\PYZus{}cycle}\PY{p}{(}\PY{n}{cycle2D}\PY{p}{(}\PY{l+m+mi}{1}\PY{p}{,} \PY{p}{[}\PY{l+m+mi}{0}\PY{p}{,} \PY{l+m+mi}{0}\PY{p}{]}\PY{p}{,} \PY{o}{\PYZhy{}}\PY{l+m+mi}{1}\PY{p}{)}\PY{p}{,} \PY{l+s+s2}{\PYZdq{}}\PY{l+s+s2}{a}\PY{l+s+s2}{\PYZdq{}}\PY{p}{)}
\end{Verbatim}
\end{tcolorbox}

    Then, we add the centre \(C\) of \(a\) as a specific point:

    \begin{tcolorbox}[breakable, size=fbox, boxrule=1pt, pad at break*=1mm,colback=cellbackground, colframe=cellborder]
\prompt{In}{incolor}{4}{\boxspacing}
\begin{Verbatim}[commandchars=\\\{\}]
\PY{n}{C}\PY{o}{=}\PY{n}{F}\PY{o}{.}\PY{n}{add\PYZus{}point}\PY{p}{(}\PY{n}{cycle2D}\PY{p}{(}\PY{n}{F}\PY{o}{.}\PY{n}{get\PYZus{}cycle}\PY{p}{(}\PY{n}{a}\PY{p}{)}\PY{p}{[}\PY{l+m+mi}{0}\PY{p}{]}\PY{p}{)}\PY{o}{.}\PY{n}{center}\PY{p}{(}\PY{p}{)}\PY{p}{,} \PY{l+s+s2}{\PYZdq{}}\PY{l+s+s2}{C}\PY{l+s+s2}{\PYZdq{}}\PY{p}{)}
\end{Verbatim}
\end{tcolorbox}

    Now, we want to add a line \(l\) tangent to \(a\). A straight line is characterised
among cycles by its orthogonality to infinity (``passes the infinity'').
The line \(l\) is not uniquely defined by these two conditions (the tangency and
orthogonality) because the point of its contact to \(a\) can be
arbitrary. Therefore its coefficients contain a free variable like
\verb|t_l| as can be seen from the figure printout.

    \begin{tcolorbox}[breakable, size=fbox, boxrule=1pt, pad at break*=1mm,colback=cellbackground, colframe=cellborder]
\prompt{In}{incolor}{5}{\boxspacing}
\begin{Verbatim}[commandchars=\\\{\}]
\PY{n}{l}\PY{o}{=}\PY{n}{symbol}\PY{p}{(}\PY{l+s+s2}{\PYZdq{}}\PY{l+s+s2}{l}\PY{l+s+s2}{\PYZdq{}}\PY{p}{)}
\PY{n}{F}\PY{o}{.}\PY{n}{add\PYZus{}cycle\PYZus{}rel}\PY{p}{(}\PY{p}{[}\PY{n}{is\PYZus{}tangent}\PY{p}{(}\PY{n}{a}\PY{p}{)}\PY{p}{,}\PY{n}{is\PYZus{}orthogonal}\PY{p}{(}\PY{n}{F}\PY{o}{.}\PY{n}{get\PYZus{}infinity}\PY{p}{(}\PY{p}{)}\PY{p}{)}\PY{p}{,}\PYZbs{}
        \PY{n}{only\PYZus{}reals}\PY{p}{(}\PY{n}{l}\PY{p}{)}\PY{p}{]}\PY{p}{,}\PY{n}{l}\PY{p}{)}
\PY{n+nb}{print}\PY{p}{(}\PY{n}{F}\PY{o}{.}\PY{n}{string}\PY{p}{(}\PY{p}{)}\PY{p}{)}
\end{Verbatim}
\end{tcolorbox}

\begin{small}
    \begin{Verbatim}[commandchars=\\\{\}]
C-(0): \{`0, [[1,0]]\textasciitilde{}C-(0), 0', -3\} --> (C);  <-- ()
C-(1): \{`0, [[0,1]]\textasciitilde{}C-(1), 0', -3\} --> (C);  <-- ()
infty: \{`0, [[0,0]]\textasciitilde{}infty, 1', -2\} --> (C,l);  <-- ()
R: \{`0, [[0,1]]\textasciitilde{}R, 0', -1\} --> ();  <-- ()
C: \{`1, [[0,0]]\textasciitilde{}C, 0', 0\} --> ();  <-- (C/o,infty|d,C-(0)|o,C-(1)|o)
a: \{`1, [[0,0]]\textasciitilde{}a, -1', 0\} --> (l);  <-- ()
l:
\{`0, [[-1/8*sqrt(8)*sqrt(2)*cos(t\_l),-1/64*sqrt(2)*sin(t\_l)*sqrt(512)]]\textasciitilde{}l,1',
`0, [[1/8*sqrt(8)*sqrt(2)*cos(t\_l),-1/64*sqrt(2)*sin(t\_l)*sqrt(512)]]\textasciitilde{}l,1',
`0, [[1/8*sqrt(8)*cos(t\_l)*sqrt(2),1/64*sin(t\_l)*sqrt(2)*sqrt(512)]]\textasciitilde{}l,1',
`0, [[-1/8*sqrt(8)*cos(t\_l)*sqrt(2),1/64*sin(t\_l)*sqrt(2)*sqrt(512)]]\textasciitilde{}l,1',1\}
 --> ();  <-- (a|t,infty|o,l|r)
Altogether 10 cycles in 7 cycle\_nodes.
    \end{Verbatim}
  \end{small}

  Note, that the parametrisation of the line \(l\) uses trigonometric
  functions in order to avoid square roots appearing in the solutions
  of the quadratic tangency relation. With such substitution automatic
  simplifications of algebraic expressions are much more efficient.

    At the next step we add the point \(P\) of contact of the circle \(a\) and the line
\(l\). A point belongs to a cycle if the point is orthogonal the cycle.
Also a point is characterised among all cycle by orthogonality to
itself. To define the latter reflexive condition we need to ``pre-cook''
its symbol in advance.

    \begin{tcolorbox}[breakable, size=fbox, boxrule=1pt, pad at break*=1mm,colback=cellbackground, colframe=cellborder]
\prompt{In}{incolor}{6}{\boxspacing}
\begin{Verbatim}[commandchars=\\\{\}]
\PY{n}{P}\PY{o}{=}\PY{n}{symbol}\PY{p}{(}\PY{l+s+s2}{\PYZdq{}}\PY{l+s+s2}{P}\PY{l+s+s2}{\PYZdq{}}\PY{p}{)}
\PY{n}{P}\PY{o}{=}\PY{n}{F}\PY{o}{.}\PY{n}{add\PYZus{}cycle\PYZus{}rel}\PY{p}{(}\PY{p}{[}\PY{n}{is\PYZus{}orthogonal}\PY{p}{(}\PY{n}{P}\PY{p}{)}\PY{p}{,} \PY{n}{is\PYZus{}orthogonal}\PY{p}{(}\PY{n}{a}\PY{p}{)}\PY{p}{,} \PYZbs{}
        \PY{n}{is\PYZus{}orthogonal}\PY{p}{(}\PY{n}{l}\PY{p}{)}\PY{p}{,} \PY{n}{only\PYZus{}reals}\PY{p}{(}\PY{n}{P}\PY{p}{)}\PY{p}{]}\PY{p}{,} \PY{n}{P}\PY{p}{)}
\end{Verbatim}
\end{tcolorbox}

    Finally we add the radius \(r\) passing \(P\): it is a straight line (is
orthogonal to the infinity) and passes both \(P\) and \(C\) (is
orthogonal to each of them).

    \begin{tcolorbox}[breakable, size=fbox, boxrule=1pt, pad at break*=1mm,colback=cellbackground, colframe=cellborder]
\prompt{In}{incolor}{7}{\boxspacing}
\begin{Verbatim}[commandchars=\\\{\}]
\PY{n}{r}\PY{o}{=}\PY{n}{F}\PY{o}{.}\PY{n}{add\PYZus{}cycle\PYZus{}rel}\PY{p}{(}\PY{p}{[}\PY{n}{is\PYZus{}orthogonal}\PY{p}{(}\PY{n}{P}\PY{p}{)}\PY{p}{,} \PY{n}{is\PYZus{}orthogonal}\PY{p}{(}\PY{n}{C}\PY{p}{)}\PY{p}{,}\PYZbs{}
         \PY{n}{is\PYZus{}orthogonal}\PY{p}{(}\PY{n}{F}\PY{o}{.}\PY{n}{get\PYZus{}infinity}\PY{p}{(}\PY{p}{)}\PY{p}{)}\PY{p}{]}\PY{p}{,} \PY{l+s+s2}{\PYZdq{}}\PY{l+s+s2}{r}\PY{l+s+s2}{\PYZdq{}}\PY{p}{)}
\end{Verbatim}
\end{tcolorbox}

Recall, that \(r\) depends on the same free parameters as \(l\). Now,
we check the orthogonality relation between \(r\) and \(l\). Because
two of the above conditions were quadratic (i.e., tangency and
self-orthogonality), each of them doubled the number of
solutions. Therefore, there are four instances of the cycle \(r\), and each
of them is checked separately.

    \begin{tcolorbox}[breakable, size=fbox, boxrule=1pt, pad at break*=1mm,colback=cellbackground, colframe=cellborder]
\prompt{In}{incolor}{8}{\boxspacing}
\begin{Verbatim}[commandchars=\\\{\}]
\PY{n}{Res}\PY{o}{=}\PY{n}{F}\PY{o}{.}\PY{n}{check\PYZus{}rel}\PY{p}{(}\PY{n}{l}\PY{p}{,}\PY{n}{r}\PY{p}{,}\PY{l+s+s2}{\PYZdq{}}\PY{l+s+s2}{orthogonal}\PY{l+s+s2}{\PYZdq{}}\PY{p}{)}
\PY{k}{for} \PY{n}{i} \PY{o+ow}{in} \PY{n+nb}{range}\PY{p}{(}\PY{n+nb}{len}\PY{p}{(}\PY{n}{Res}\PY{p}{)}\PY{p}{)}\PY{p}{:}
    \PY{n+nb}{print}\PY{p}{(}\PY{l+s+s2}{\PYZdq{}}\PY{l+s+s2}{Tangent and radius are orthogonal: }\PY{l+s+si}{\PYZpc{}s}\PY{l+s+s2}{\PYZdq{}} \PY{o}{\PYZpc{}}\PYZbs{}
    \PY{n+nb}{bool}\PY{p}{(}\PY{n}{Res}\PY{p}{[}\PY{n}{i}\PY{p}{]}\PY{o}{.}\PY{n}{subs}\PY{p}{(}\PY{n+nb}{pow}\PY{p}{(}\PY{n}{cos}\PY{p}{(}\PY{n}{wild}\PY{p}{(}\PY{l+m+mi}{0}\PY{p}{)}\PY{p}{)}\PY{p}{,}\PY{l+m+mi}{2}\PY{p}{)}\PY{o}{==}\PY{l+m+mi}{1}\PY{o}{\PYZhy{}}\PY{n+nb}{pow}\PY{p}{(}\PY{n}{sin}\PY{p}{(}\PY{n}{wild}\PY{p}{(}\PY{l+m+mi}{0}\PY{p}{)}\PY{p}{)}\PY{p}{,}\PY{l+m+mi}{2}\PY{p}{)}\PY{p}{)}\PYZbs{}
        \PY{o}{.}\PY{n}{normal}\PY{p}{(}\PY{p}{)}\PY{p}{)}\PY{p}{)}
\end{Verbatim}
\end{tcolorbox}

    \begin{Verbatim}[commandchars=\\\{\}]
Tangent and radius are orthogonal: True
Tangent and radius are orthogonal: True
Tangent and radius are orthogonal: True
Tangent and radius are orthogonal: True
    \end{Verbatim}

    Note, that we had uses the Pythagoras substitution
    \(\cos^2 t = 1 - \sin^2 t\) to assist the CAS with algebraic
    simplifications. Another interesting moment is that the entire
    construction is based on the single relation of orthogonality
    (apart from the tangency required to define \(l\)).

Many more examples of computer-assisted proofs were presented
in~\cite{Kisil12a} and are freely available for evaluation
at~\cites{Kisil19c,Kisil19d}.

\section{Software Impact}
\label{sec:impl-empir-results}



\begin{figure}[htbp]
  \centering
  \includegraphics[width=.47\textwidth]{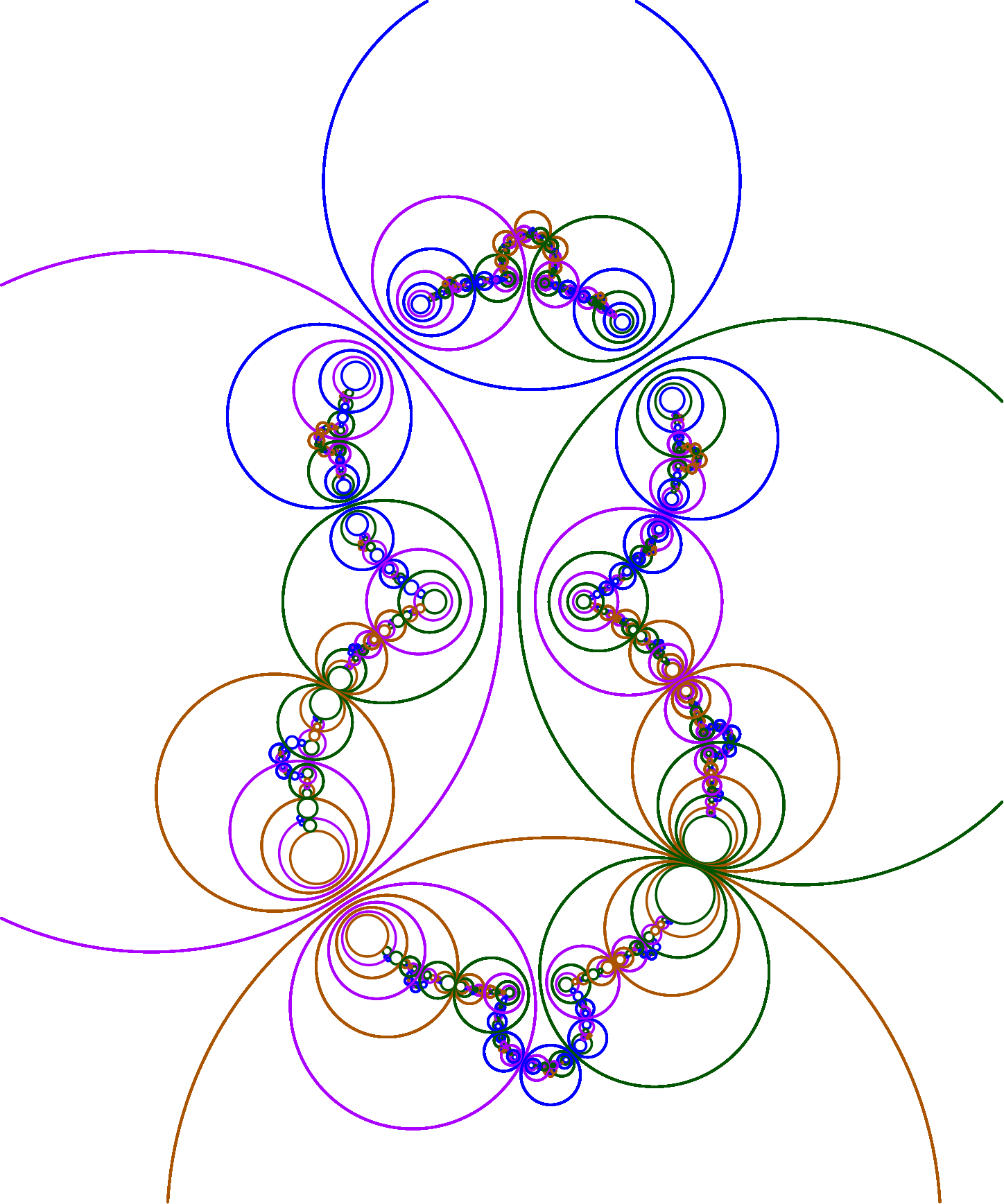}
  \includegraphics[width=.47\textwidth]{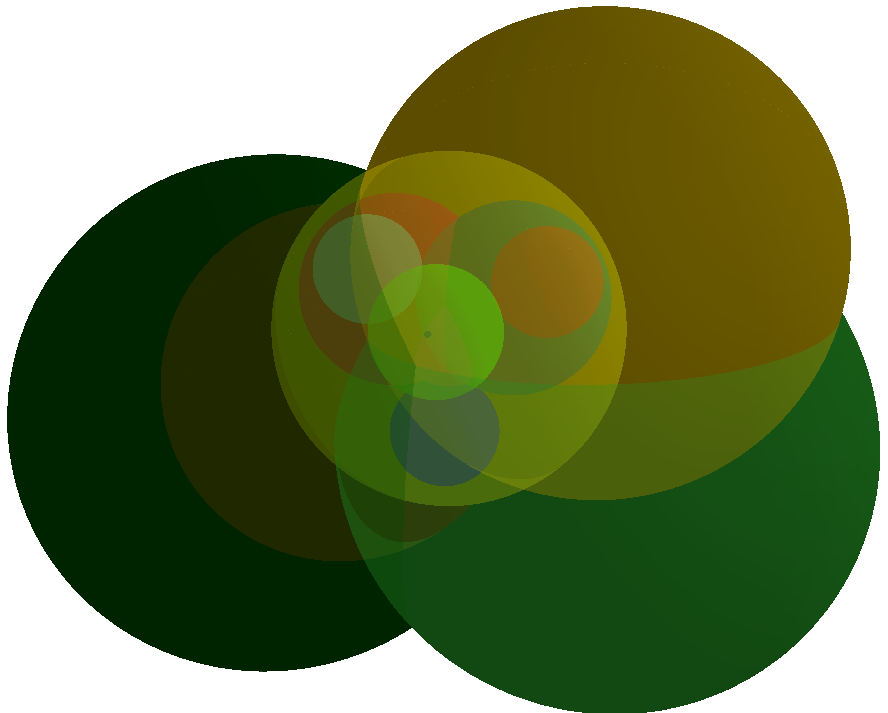}
  \caption{Left: a Kleinian group generated by four cycle reflections.
    Right: an example of Apollonius problem in three dimensions.}
  \label{fig:apollonius-3D}
\end{figure}

The first library \textsf{cycle} was initially reported in~\cite{Kisil05b} and
since then was employed as the main research tool in several papers~\cite{Kisil05a,Kisil06a} and  monograph~\cite{Kisil12a}. New
results, e.g. elliptic and hyperbolic conformal version of the
nine-point theorem, discovered with the second library \textsf{figure}
were published in~\cite{Kisil14b,Kisil19a}, cf. Fig.~\ref{fig:illustr-conf-nine}. 
The package has several different uses:
\begin{itemize}
\item It is easy to produce high-quality illustrations, which are
  fully-accurate with automatic evaluation of cycles' parameters, cf. Fig.~\ref{fig:apollonius-3D}
\item The package can be used for computer experiments in M\"obius--Lie
  geometry. 
\item Since the library is based on the \GiNaC\ system, which provides a
  symbolic computation engine, there is a possibility to make fully
  automatic proofs of various statements in M\"obius--Lie geometry, see the previous Section. 
\item Last but not least, the combination of classical beauty of Lie
  sphere geometry and modern computer technologies is a useful
  pedagogical tool to widen interest in mathematics through visual and
  hands-on experience. A dedicated GUI makes it accessible for all
  schoolchildren. Several videos created with the package are uploaded
  to the popular video hosting~\cite{Kisil16a} to disseminate
  aesthetical attraction of mathematics to wider public.
\end{itemize}


As a non-trivial example of automated proof accomplished by the \texttt{figure}
library for the first time, we present a FLT-invariant version of the
classical nine-point theorem~\cites{Kisil14b,Kisil19a}, which is
illustrated in the attached video.

\section{Conclusions}
\label{sec:conclusions}


The package accumulates more than 15 years of work and is still
under active development. The design of the library \texttt{figure}
shaped the general theoretical approach to the extension of M\"obius--Lie geometry~\cites{Kisil14b,Kisil19a}, which leaded to specific
realisations
in~\cites{Kisil15a,Kisil14a,KisilReid18a}. Furthermore, it
shall be helpful for computer experiments in Lie sphere geometry of
indefinite or nilpotent metrics since our intuition is not elaborated
there in contrast to the Euclidean
space~\cites{Kisil06a,Kisil05a,Kisil07a}. Some advances in the
two-dimensional space were achieved
recently~\cites{Kisil12a,Mustafa17a}, however further developments in
higher dimensions are still awaiting their researchers.

The current version (v3.5.7) of the package was tested on major
desktop platforms: Linux (Debian, Ubuntu and CentOS distributions),
Windows\,10, Mac\,OS\,X. Pre-compiled binaries for these OSes are
provided on the project Web page~\cite{Kisil14b}.  The interactive
\Python/\Jupyter\ shells can be used from cloud
services CoLab~\cite{Kisil19d} and CodeOcean~\cite{Kisil19c}.

\section*{Acknowledgements}
\label{sec:acknowledgements}


I am grateful to Prof.~Jay~P.~Fillmore for stimulating discussion,
which enriched the package.  The University of Leeds provided supports
for several student projects, which helped to develop the
package. L.~Hutton wrote the initial version of the GUI and helped to
port the package to Mac\,OS\,X.









\clearpage
\section*{Required Metadata}
\label{}

\section*{Current code version}
\label{}


\begin{table}[!h]
\begin{tabular}{|l|p{.44\textwidth}|p{.44\textwidth}|}
\hline
\textbf{Nr.} & \textbf{Code metadata description} & \textbf{Please fill in this column} \\
\hline
C1 & Current code version & v3.5.7 \\
\hline
  C2 & Permanent link to code/repository used of this code version &
   \url{https://sourceforge.net/projects/moebinv/files/releases/moebinv_3.5.7.orig.tar.gz}
  \\
\hline
C3 & Legal Code License   & GPL v3.0 \\
\hline
C4 & Code versioning system used & git \\
\hline
C5 & Software code languages, tools, and services used & \CPP, \Python \\
\hline
C6 & Compilation requirements, operating environments \& dependencies & Compiled for Linux, Win32, Mac OS\,X using \href{https://gcc.gnu.org}{GNU g++} or \href{https://clang.llvm.org/}{clang}.\\
\hline
C7 & If available Link to developer documentation/manual &                                                             
\url{https://sourceforge.net/projects/moebinv/files/docs/figure.pdf} \\
\hline
  C8 & Support email for questions & \href{mailto:kisilv@maths.leeds.ac.uk}{kisilv@maths.leeds.ac.uk}
  \href{mailto:V.Kisil@leeds.ac.uk}{V.Kisil@leeds.ac.uk}\\
\hline
\end{tabular}
\caption{Code metadata (mandatory)}
\label{} 
\end{table}




\small
\providecommand{\noopsort}[1]{} \providecommand{\printfirst}[2]{#1}
  \providecommand{\singleletter}[1]{#1} \providecommand{\switchargs}[2]{#2#1}
  \providecommand{\irm}{\textup{I}} \providecommand{\iirm}{\textup{II}}
  \providecommand{\vrm}{\textup{V}} \providecommand{\cprime}{'}
  \providecommand{\eprint}[2]{\texttt{#2}}
  \providecommand{\myeprint}[2]{\texttt{#2}}
  \providecommand{\arXiv}[1]{\myeprint{http://arXiv.org/abs/#1}{arXiv:#1}}
  \providecommand{\doi}[1]{\href{http://dx.doi.org/#1}{doi:
  #1}}\providecommand{\CPP}{\texttt{C++}}
  \providecommand{\NoWEB}{\texttt{noweb}}
  \providecommand{\MetaPost}{\texttt{Meta}\-\texttt{Post}}
  \providecommand{\GiNaC}{\textsf{GiNaC}}
  \providecommand{\pyGiNaC}{\textsf{pyGiNaC}}
  \providecommand{\Asymptote}{\texttt{Asymptote}}

\end{document}